# Magnetic and electrical transport anomalies of RMAs$_2$ (R= Pr and Sm, M= Ag and Au)

K. Mukherjee and E.V. Sampathkumaran
Tata Institute of Fundamental Research, Homi Bhabha Road, Colaba, Mumbai-400005, India
*and*
D. Rutzinger, Th. Doert, M. Ruck
Department of Chemistry and Food Chemistry, Technische Universität Dresden, D-01062 Dresden, Germany.

**Abstract**
The results of magnetization, heat-capacity and electrical resistivity ($\rho$) studies of the compounds, RMAs$_2$ (R= Pr and Sm; M= Ag, Au), crystallizing in HfCuSi$_2$-derived structure are reported. PrAgAs$_2$ orders antiferromagnetically at $T_N$ = 5 K. The Au analogue, however, does not exhibit long range magnetic order down to 1.8 K. We infer that this is due to subtle differences in their crystallographic features, particularly noting that both the Sm compounds with identical crystal structure as that of former order magnetically nearly at the same temperature (about 17 K). It appears that, in PrAgAs$_2$, SmAgAs$_2$, and SmAuAs$_2$, there is an additional magnetic transition at a lower temperature, as though the similarity in the crystal structure results in similarities in magnetism as well. The $\rho$ for PrAgAs$_2$ and PrAuAs$_2$ exhibits negative temperature coefficient in some temperature range in the paramagnetic state. SmAuAs$_2$ exhibits magnetic Brillouin-zone gap effect in $\rho$ at $T_N$, while SmAgAs$_2$ shows a well-defined broad minimum well above $T_N$ around 45 K. Thus, these compounds reveal interesting magnetic and transport properties.

## I. Introduction

Following the discovery of superconductivity in the arsenide family in recent years, there have been explosive activities on arsenic containing transition-metal compounds. In particular, the ThCr$_2$Si$_2$-derived tetragonal compounds have attracted a lot of interest in this respect. The arsenic compounds with 1:1:2 composition, e.g., HfCuSi$_2$-type layered structure [1-3] – a defective variant of the ThCr$_2$Si$_2$ structure – however, has not been explored sufficiently for magnetic and superconducting anomalies, barring a few reports on Cu-based compounds RCuAs$_2$ (R = rare earth metals) [4-8]. Though corresponding Ag and Au-based compounds were also identified [8-10], very little work has been reported on these compounds. Considering that the Cu-based compounds, even for those rare-earths with 4f-stability, have been reported to exhibit interesting properties, we considered it worthwhile to probe these Ag and Au containing compounds as well. We believe that the knowledge thus gathered would eventually contribute to global understanding of the properties of the arsenide family. Among these Ag and Au compounds, interesting Kondo anomalies for Ce analogues have been reported in the recent literature [7]. Here we report the results of our investigation on the compounds PrMAs$_2$ and SmMAs$_2$ with M = Ag and Au. We find that the temperature ($T$) dependencies of the electrical resistivity ($\rho$) of these compounds are interesting and not-so-commonly known for Pr and Sm compounds.

## II. Crystallographic features

The crystal structures of these ternary compounds can be understood as variants of the tetragonal HfCuSi$_2$ type. In this structure, PbO-like layers of the coinage metals and arsenic are stacked along [001] with planar As sheets, separated by layers of the respective rare earth metals (figure 1). The undistorted HfCuSi$_2$ structure comprises of As square layers. Due to a Peierls-like distortion, two different structure variants are found for the title compounds: PrAgAs$_2$, SmAgAs$_2$ and SmAuAs$_2$ crystallize as two-fold superstructures in orthorhombic space group *Pmcn* (no. 62) with the As atoms of their planar layers forming zigzag chains (figure 2), whereas PrAuAs$_2$ adopt a fourfold superstructure (orthorhombic space group *Pmca*) with cis-trans chains of As atoms (figure 2) [10].

## III. Experimental details

The sample preparations were carried out in an argon-filled glove box (M. Braun, *p* (O$_2$) ≤ 1 ppm, *p* (H$_2$O) ≤ 1 ppm) with purification of argon with molecular sieve and copper catalyst. Pieces of praseodymium (>99.9% purity, Treibacher AG) or samarium (99.9%, Chempur GmbH), freshly filed from rods of the respective rare earth metals, silver (powder, 99.9%, Chempur GmbH) or gold (powder, >99.9% purity, Chempur GmbH), and arsenic (powder, > 99.997 % metal based, Aldrich; As$_2$O$_3$ removed by sublimation prior to use) were mixed in the atomic ratio of 1:1:2. The reactions were carried out in a six-fold excess of a LiCl/KCl flux (LiCl, KCl: powder, p. a. , Merck, dried at 410 K in dynamic vacuum prior to use) in carbon crucibles which were sealed in evacuated silica ampoules. The ampoules were heated up to 1023 K for 48 hours, annealed for 96 hours, and cooled to 623 K over a period of 192 hours. The flux was removed with water and the polycrystalline products were washed with ethanol. The specimens thus obtained were shiny black platelets and stable in air. Pellets with 8 mm diameter and approximately 2 mm height were obtained at ambient temperature by pressing poly crystals obtained by crushing under argon. The pellets were then sintered at 523 K for 24 h in evacuated sealed tubes.

X-ray powder diffraction patterns of the reaction products were recorded in order to check the sample purity. The measurements were performed in transmission geometry on a Stadi P diffractometer (Stoe & Cie., Darmstadt, Germany) equipped with an IP-PSD using Ge monochromatized Cu *K*α$_1$ radiation. The evaluation of the patterns was done with the WinXPow program package [11]. All reflections can unambiguously be indexed (Fig. 3) with respect to theoretical patterns which were calculated on the basis of the structure models obtained from single crystal data [10]. No lines of impurity phases were detected.

Dc magnetization (*M*) measurements (1.8-300 K) were performed with the help of a commercial SQUID magnetometer (Quantum Design). The ρ measurements (1.8-300 K) in zero as well as in the presence of magnetic fields (*H*) were carried out employing a commercial physical properties measurements system (Quantum Design) and heat-capacity (*C*) data were also collected with the same instrument by a relaxation method.

### IV.     Results and discussion

We show the results for PrAgAs$_2$ in figures 4 and 5. The magnetic susceptibility (χ) obtained in a field of 5 kOe exhibits Curie-Weiss behavior in the range 30-300K (figure 4a) and there is a deviation from this behavior at low temperatures (<30 K) which is usually attributed to crystal-field effects in the literature. The effective moment (μ$_{eff}$ ~3.6 μ$_B$) obtained from the linear region confirms trivalency of Pr. The paramagnetic Curie temperature (θ$_p$) is found to be ~ –7 K and the negative sign indicates dominance of antiferromagnetic interactions. In order to understand the low temperature behavior, we show the plot of *M*/*H* as a function of *T* measured in a field of 100 Oe in figure 4b. In this figure, there is a distinct peak at 5 K. Below 4 K, there is a bifurcation of the curves obtained under zero-field-cooled (ZFC) and field-cooled (FC) conditions (from 50 K) of the specimen. C(*T*) plot (figure 4c) reveals a prominent λ-anomaly near 5 K establishing long-range magnetic order. Therefore, the bifurcation of ZFC-FC curves mentioned above is not due to spin-glass freezing. This conclusion was further confirmed by the absence of frequency dependence of ac susceptibility. Isothermal magnetization at 1.8 K exhibits a sharp increase for initial applications of field (<5 kOe), followed by a sluggish variation at higher fields (figure 4d) and there is a weak irreversibility at low fields. This *M*(*H*) behavior implies that there is a ferromagnetic component as well and therefore this compound could be classified as a canted antiferromagnet. However, it appears that there is a subtle change in the antiferromagnetic structure at 4 K, as evidenced by the differences in the *M*(*H*) feature in the low-field range above and below 4 K. That is, at 1.8 K, there is a step in the virgin curve at low fields, whereas at 4.5 K, this step is absent (see inset of figure 4d). In addition, the *M*(*H*) plot at 4.5 K is not hysteretic. A careful look at the derivative of ρ (see inset of figure 5a) also offers a support to the existence two magnetic transitions in the close vicinity of 4 K, apart from the fact that the transport behavior is overall quite fascinating (figure 5a). ρ increases with decreasing temperature exhibiting a maximum around 100 K and a minimum around 40 K followed by an upturn with a further decrease of temperature down to lowest measured temperature. It appears that this is another example for the family of compounds exhibiting magnetic precursor effects [12]. The origin of negative temperature coefficient of ρ in the paramagnetic state above 100 K is not clear and future studies should focus on whether this is due to Pr 4f hybridization effects. We have also extended low-field (100 Oe) magnetization studies up to 250 kOe to explore whether there is any other magnetic anomaly around 100 K, but we couldn't detect any. With respect to the behavior near *T$_N$*, it should be noted that there is no fall of ρ below *T$_N$*, and the upturn keeps continuing down to lowest measured temperature. This establishes that, at *T$_N$*, there is a magnetic pseudogap formation. This is further established by the observation that an application of magnetic field gradually decreases ρ (figure 5b) resulting in negative magnetoresistance (MR= [ρ(H)-ρ(0)/ρ(0)]), which is prominent below *T$_N$* only (compare in-field and zero-field curves figure 5a). Despite the fact that there is no fall due to loss of spin-disorder contribution at the onset of magnetic transition, the inset of figure 5 reveals that there are distinct changes in the slopes of ρ(*T*) at 4 and 5 K, thereby revealing that the ZFC-FC bifurcation of χ(*T*) curves at 4 K in figure 4b must have its origin in a subtle change in the magnetic structure (supporting the inference from *M*(*H*) curve in the low-field range, made above). These results overall establish that this compound is an antiferromagnet with interesting transport anomalies, even in the paramagnetic state.

With respect to PrAuAs$_2$, $\chi$ monotonically increases with decreasing $T$, exhibiting Curie-Weiss behavior down to 20 K (Fig. 6a) typical of trivalent Pr ions. The sign and the magnitude of $\theta_p$ is the same as in PrAgAs$_2$. There is no difference in low-field ZFC and FC curves down to 2 K (Fig. 6b) and $\chi$ continues to rise down to 1.8 K. These features indicate that there is no long range magnetic order down to 1.8 K. This is consistent with the absence of any λ-anomaly in $C$ ($T$) (figure 6c). However, a plot of C/$T$ reveals a gradual fall below 7 K. In addition, isothermal $M$ at 1.8 K shows a marginal deviation at high fields from the low-field linear behavior (see inset in Fig. 6a). In order to understand these features, we have taken MR data as a function of $H$ (figure 7a). It is found that MR (with negative sign) varies quadratically with $H$ initially typical of paramagnets above 10 K, as shown for 20 K in figure 7a. However, as the temperature is lowered below 7 K, the variation of MR is steeper (with the negative sign of MR), as though there is a magnetic ordering. Spin-glass behavior is ruled out considering absence of a bifurcation of ZFC and FC $\chi$ curves. A way to reconcile this behavior is to propose that there are short range correlations developing gradually below 7 K in this compound. In fact, the temperature derivative of $\chi$ shows a significant change as shown in figure 6b around this temperature which appears to endorse this inference. In figure 7b, we show $\rho(T)$ behavior in zero field and in 50 kOe. While the slope of $\rho$ is positive above about 150 K, there is a broad, but a distinct, minimum around 100 K with the upturn persisting down to 2 K. These features persist even for $H$= 50 kOe. If one assumes that long-range magnetic ordering sets in below 1.8 K, then the low temperature increase could be of the same origin as in the Ag analogue. Experiments at low temperatures (< 1.8K) are warranted for this compound to understand it better. We believe that the suppression of long range magnetic ordering in PrAuAs$_2$ with respect to Ag analogue, despite similar values of $\theta_p$, is in some way related to the fourfold superstructure with cis-trans chains of As atoms (possibly inducing magnetic frustration in some fashion). This inference gains further support from the similarities in the magnetism of Sm compounds.

The results on SmAgAs$_2$ are shown in figures 8 and 9. It is a well-known fact that Sm in its trivalent state exhibits complex temperature dependent behavior due to narrow multiplet widths and crystal-field splitting, as a result of which Curie-Weiss behavior in the paramagnetic state is unexpected. Our aim here is to focus on the low temperature behavior in the vicinity of magnetic transition. It is obvious from figure 8a that there is a peak in $\chi$(T) at ~16 K due to onset of magnetic ordering and another upturn below 8 K. This is observed irrespective of whether the specimen was cooled in zero field or in a field. While there is a prominent λ-anomaly in C near 16 K (figure 8b) establishing long range magnetic ordering at this temperature, the feature near 8 K (marked by an arrow) is however weak. Possibly, the entropy associated with the 8 K-transition is negligible. It is however important to ensure that this transition is not due to an impurity. The $\rho$(T) behavior in this regard is quite helpful to resolve this. In figure 8c, in addition to a fall below 16 K due to a loss of spin-disorder contribution, there is an upturn below 8 K. This 8K-upturn in $\rho$ can not be due to any impurity; if the 8K-feature in $\chi$ and C is attributed to impurity, the positive temperature coefficient of the main magnetic phase should have resulted in a continuous fall of $\rho$ below 8 K as well. A semiconducting impurity phase can not dominate transport behavior when its fraction is small (if inferred from the strength of the C feature). Therefore, we tend to believe that there is another magnetic feature coming from SmAgAs$_2$ phase only and the 8 K-upturn may arise from magnetism-induced pseudo-gap setting in at this temperature due to possible spin-reorientation. The change in spin alignment must be a subtle one, as there is no dramatic difference in the nature of the $M(H)$ curves above and below 8 K (compare the curves in figure 9a for these temperatures). $M$ increases in a sluggish manner with $H$ without any hysteresis. From this, we infer that the magnetic ordering is of an antiferromagnetic type in this compound. We also wanted to see whether MR can throw some light on this conclusion, but MR appears to be dominated by conduction electron contribution as indicated by its quadratic field-dependence with a positive sign. The most interesting observation for this compound is that $\rho(T)$ exhibits a minimum around 50 K, which is three times that of $T_N$ (~16 K), which could not be suppressed by an application of a magnetic field of 50 kOe (figure 8c).

In the case of SmAuAs$_2$, the features attributable to two magnetic transitions are visible in $\chi(T)$ (see figure 10) in the form of flattening around ($T_N$=) 17 K, followed by an upturn below about 12 K. There is no difference between ZFC and FC curves. The fact that these are bulk magnetic transitions is confirmed by prominent peaks in C($T$) at these temperatures (figure 10b). There is an upturn in $\rho(T)$ below 17 K (figure 10c) due to magnetic Brillouin-zone formation, thereby indicating that the magnetic transition at this temperature is of an antiferromagnetic type. However, this gap effect appears to diminish as soon as the second transition sets in, as indicated by a gradual fall of $\rho$ around this temperature. Interestingly, further lowering of temperature below 6 K results in negative temperature coefficient of $\rho$. These features are not altered in a field of 50 kOe. Thus, it appears that there are interesting changes in the Fermi surface with varying temperature below $T_N$. It is to be noted that, below $T_N$, M(H) curves are non-hysteretic varying gradually with $H$ without any tendency for saturation (figure 11a) similar to that for SmAgAs$_2$. We would like to add that the sign of MR is positive in the entire temperature range below $T_N$ (see figure 11b) varying nearly quadratically with $H$ similar to SmAgAs$_2$. With respect to the transport behavior in the paramagnetic state, unlike in SmAgAs$_2$, $\rho$ is metallic-like without any minimum.

**V.   Summary**

The magnetic and transport properties of ternary arsenides, RMAs$_2$ (R = Pr and Sm, M = Ag and Au) have been investigated. All, except PrAuAs$_2$, order antiferromagnetically at low temperatures with complex electrical resistivity behavior. It is intriguing to note that PrAgAs$_2$ and the Sm compounds are characterized by similar magnetic anomalies in the sense that there are two magnetic transitions, as though similarity in crystal structure determines this magnetic behavior. Since the onset of magnetic transition occurs at nearly the same temperature for both the Sm compounds with the same crystallographic features, a suppression of magnetic ordering in the PrAuAs$_2$ in comparison with PrAgAs$_2$ (despite the same value of $\theta_p$) may be attributable to subtle structural differences, possibly in the superstructure features as outlined in Section II, between these two Pr compounds. A point being stressed is that, even in the paramagnetic state, the transport behavior is interesting in the sense that negative temperature coefficient of $\rho(T)$ far above $T_N$ is observed in all cases except SmAuAs$_2$, with insensitivity to application of a magnetic field. Similar behavior was reported for RCuAs$_2$ (Sampathkumaran et al, Ref.12). Such a feature prior to an onset of long range magnetic order is of theoretical interest [13, 14], and it is possible to explain its presence in some ferromagnetic systems [13]. Free electron scattering on collective excitations from crystal-field levels [14] was proposed as an explanation in all magnetic materials. It is however not clear whether insensitivity of the $\rho(T)$ minimum to applications of magnetic fields can be explainable within this theory. In addition, among the two Sm systems in the same family, only one member exhibits this anomaly. It therefore appears that more theoretical work is required to address this transport anomaly.

We thank Sitikantha D Das and Kartik K Iyer for their help while carrying out experiments.

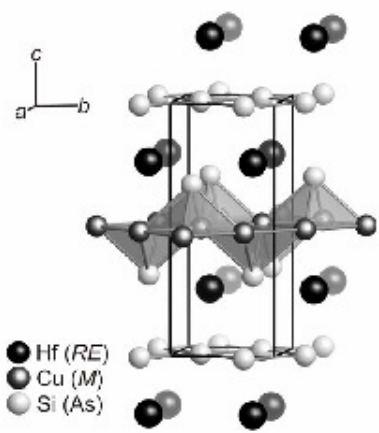

Figure 1:	Crystal structure of HfCuSi$_2$.

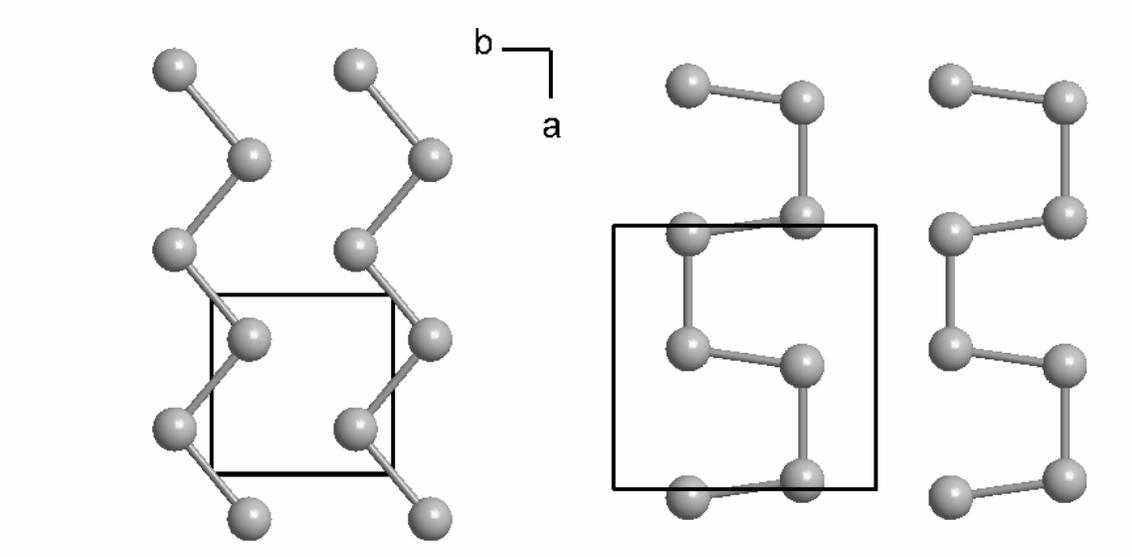

Figure 2: As-layers of the compounds crystallizing as twofold superstructures of the HfCuSi$_2$ type in *Pmcn* (left), and as fourfold superstructures in *Pmca* (right).

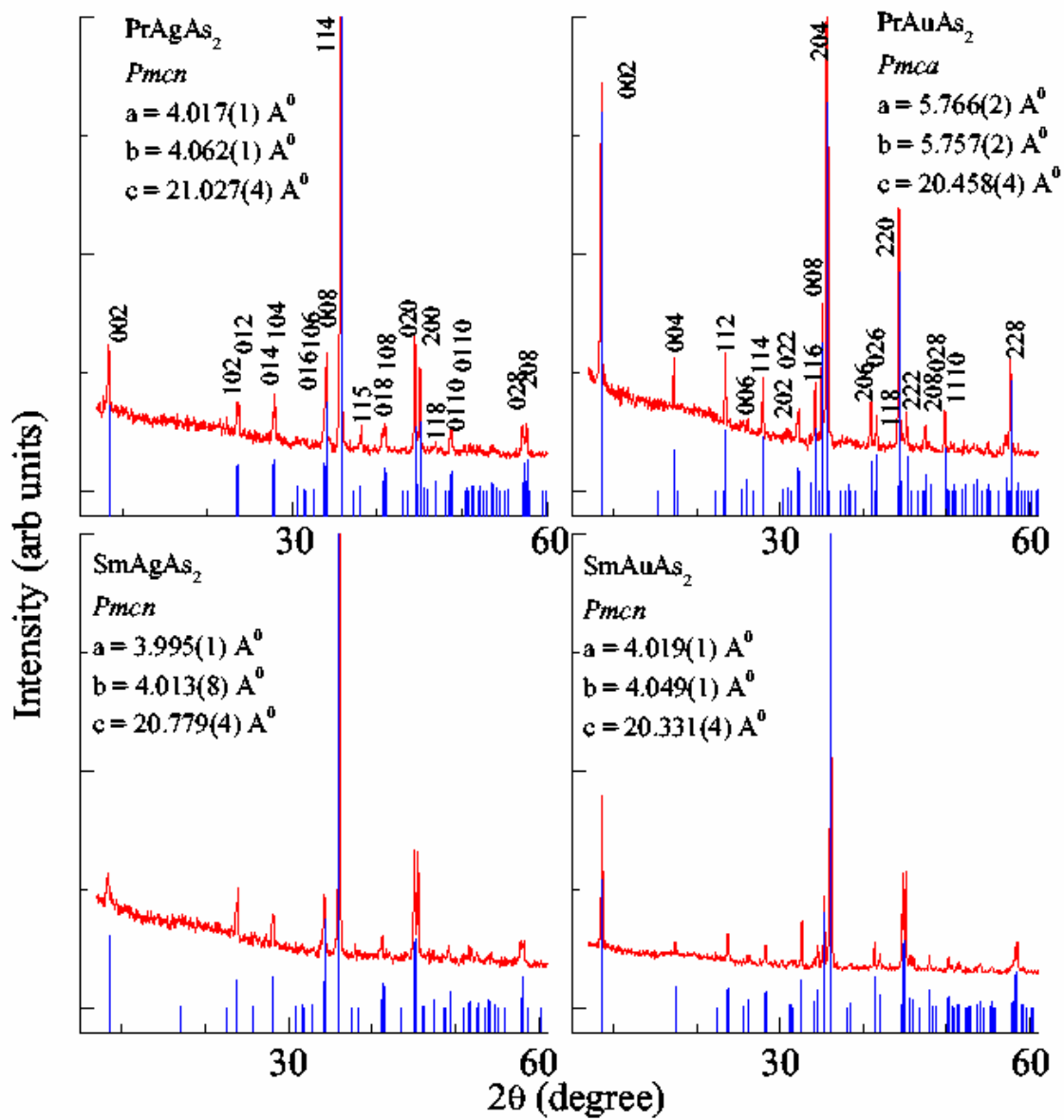

Figure 3:
X-ray powder diffraction patterns of RMAs$_2$ (R= Pr and Sm, M= Ag and Au) obtained experimentally (curves) and by calculations (vertical bars) as mentioned in text. The lattice constants, *a*, *b*, and *c*, are also included.

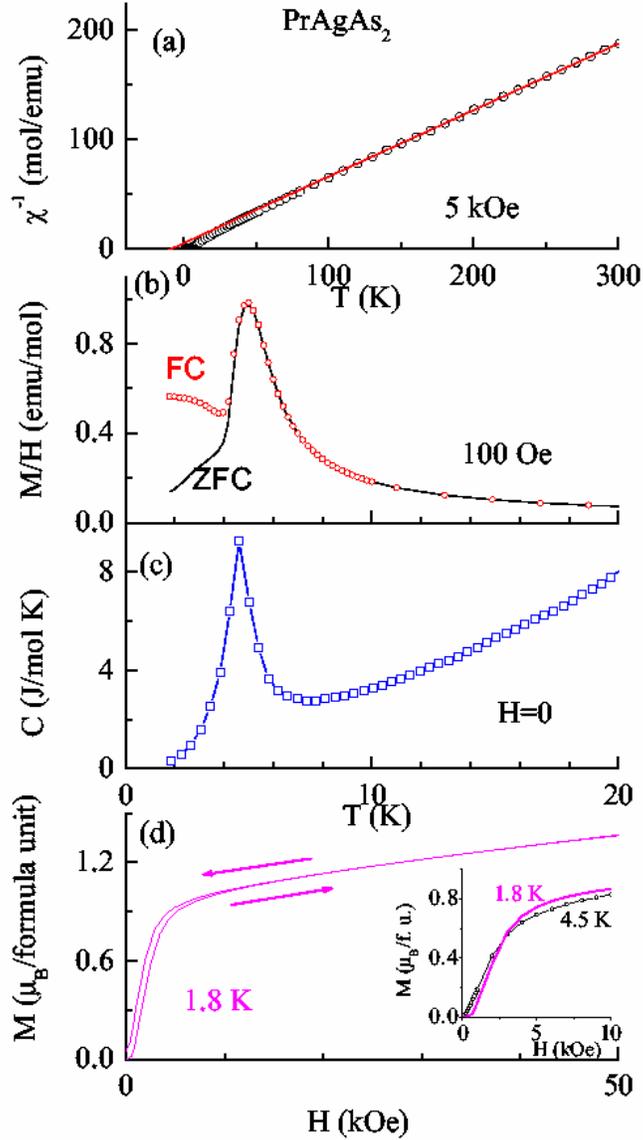

Figure 4: (**a**) Inverse magnetic susceptibility measured in a field of 5 kOe, (**b**) zero-field-cooled and field-cooled low-field magnetization, (**c**) heat capacity as a function of temperature and (**d**) isothermal magnetization as a function of magnetic field for 1.8 K is plotted for PrAgAs$_2$. In the inset of (**d**), $M(H)$ behavior with increasing field at 1.8 and 4.5 K below 10 kOe are compared. In (**a**), a line is drawn through the Curie-Weiss region. Otherwise, the lines through the data points serve as guides to the eyes.

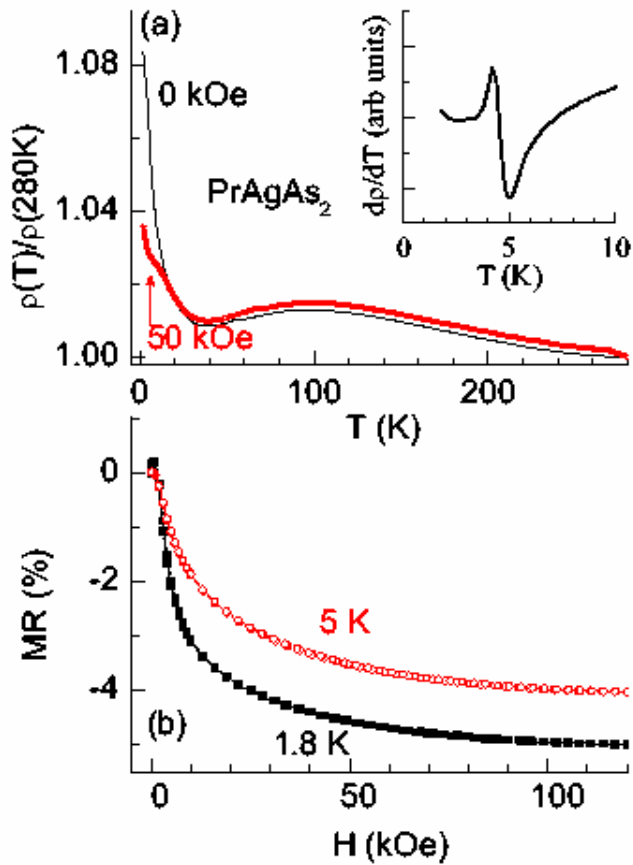

Figure 5: (**a**) Normalised electrical resistivity in zero and in 50 kOe as a function of temperature for PrAgAs$_2$. (**b**) For the same sample, the magnetoresistance as a function of magnetic field at 1.8 and 5 K is plotted. The lines through the data points serve as guides to the eyes. In the inset of (**a**), the temperature derivative of electrical resistivity is plotted to highlight the changes in slopes at 4 and 5 K.

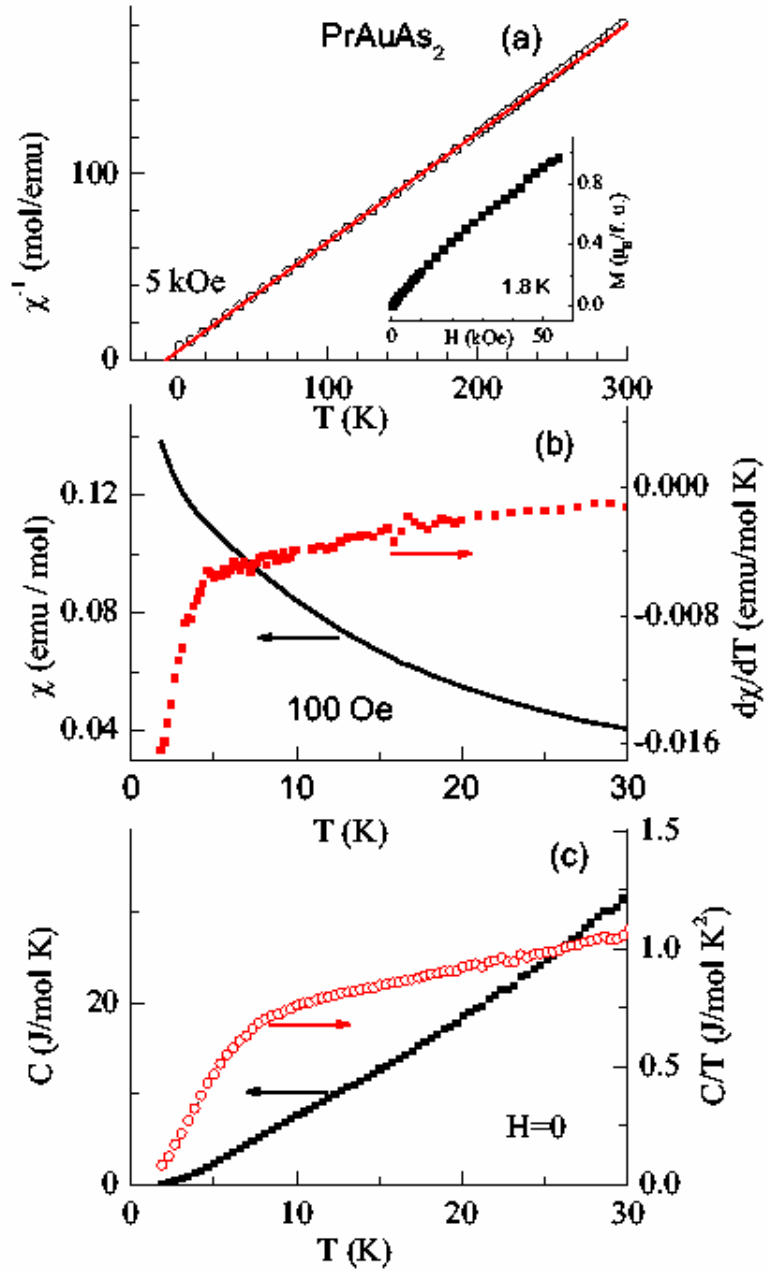

Figure 6: (**a**) Inverse magnetic susceptibility ($\chi$) measured in a field of 5 kOe, (**b**) $\chi$ and $d\chi/dT$ as a function of temperature measured in a field of 100 Oe, and (**c**) heat-capacity data as a function of temperature for $PrAuAs_2$. In the inset of (**a**), isothermal magnetization at 1.8 K is plotted. A continuous line through the Curie-Weiss region is drawn in (**a**).

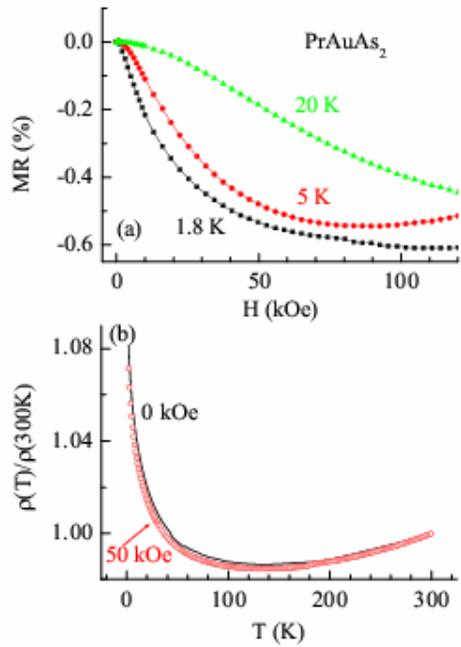

Figure 7: (**a**) Magnetoresistance as a function of magnetoresistance at 1.8, 5 and 20 K, and (**b**) normalized electrical resistivity as a function of temperature in zero and in 50 kOe for PrAuAs$_2$. Continuous lines through the data points are drawn as guides to the eyes.

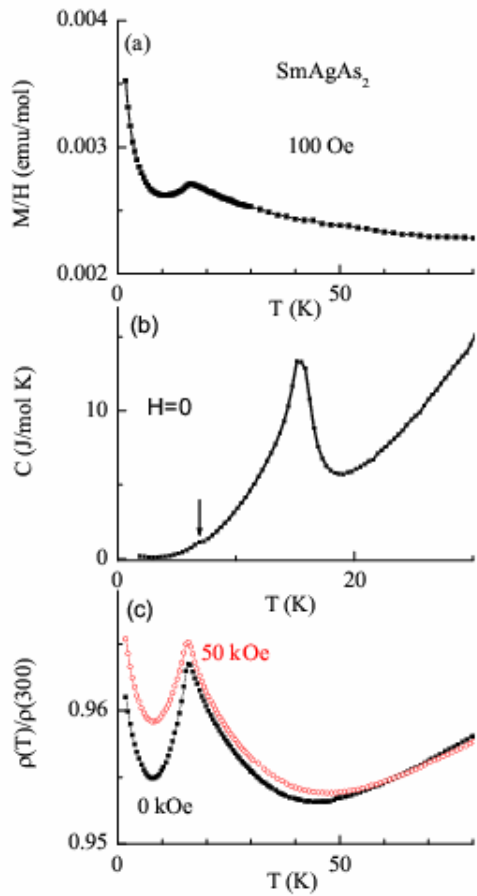

Figure 8: Temperature dependence of (**a**) magnetization, (**b**) heat-capacity and (**c**) normalized electrical resistivity (in zero field and in 50 kOe) in the vicinity of magnetic transition, for SmAgAs$_2$. The lines through the data points serve as guides to the eyes.

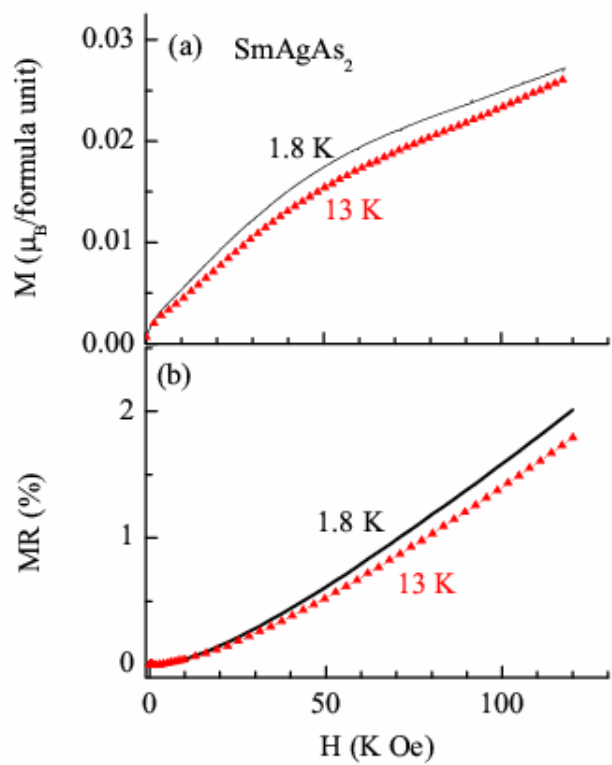

Figure 9: (**a**) Isothermal magnetization and (**b**) magnetoresistance as a function of magnetic field at 1.8 and 13 K for SmAgAs$_2$.

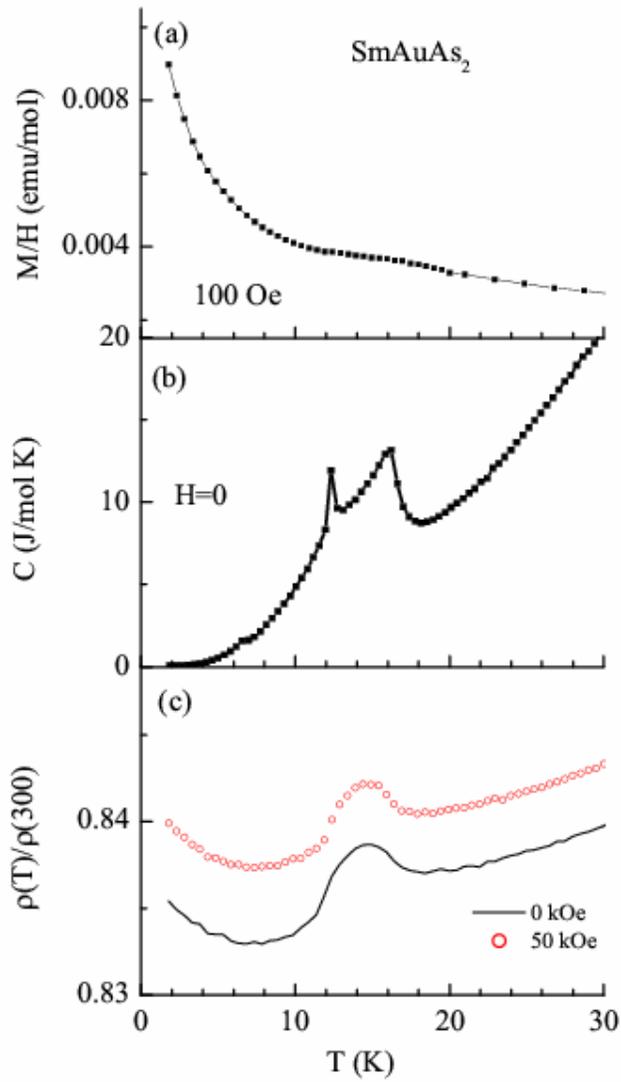

Figure 10: (**a**) Low-field magnetic susceptibility, (**b**) heat-capacity, and (**c**) normalized electrical resistivity as a function of temperature for SmAuAs$_2$.

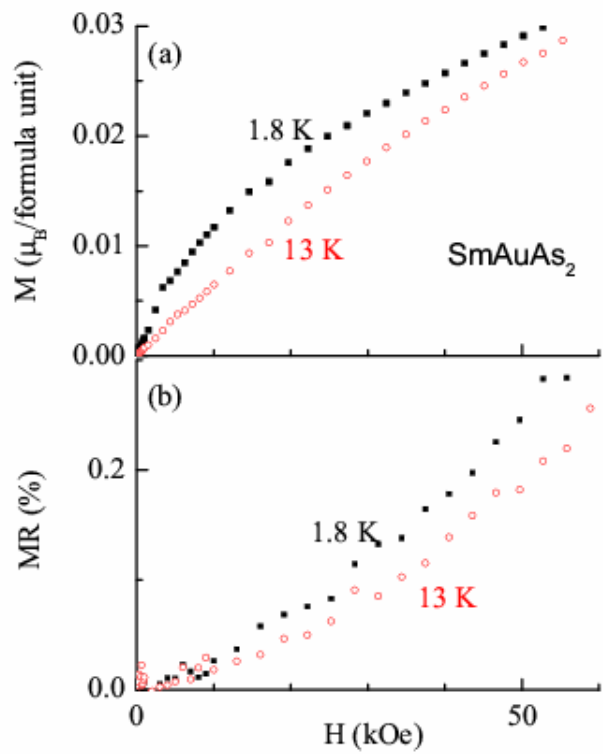

Figure 11: (**a**) Isothermal magnetization and (**b**) magnetoresistance at 1.8 and 13 K for SmAuAs$_2$.